\documentclass[12pt]{iopart}
\usepackage{iopams}
\usepackage{appendix}
\usepackage{xcolor}
\usepackage{graphicx}

\begin{document}

\title{Fluorescence detection at the atom shot noise limit for atom interferometry} 
\author{E. Rocco, R. N. Palmer, T. Valenzuela, V.  Boyer, A. Freise and K. Bongs}
\ead{e.rocco@bham.ac.uk}
\address{The School of Physics and Astronomy, University of Birmingham, Edgbaston, Birmingham B15 2TT, United Kingdom} 
\begin{abstract}
Atom interferometers are promising tools for precision measurement with applications ranging from geophysical exploration to tests of the equivalence principle of general relativity, or the detection of gravitational waves. Their optimal sensitivity is ultimately limited by their detection noise. We review resonant and near-resonant methods to detect the atom number of the interferometer outputs and we theoretically analyze the relative influence of various scheme dependent noise sources and the technical challenges affecting the detection. We show that for the typical conditions under which an atom interferometer operates, simultaneous  fluorescence detection with a CCD sensor is the optimal imaging scheme. We extract the laser beam parameters such as detuning, intensity, and duration, required for reaching the atom shot noise limit.
\end{abstract}
 
\pacs{67.85.-d, 03.75.Be, 03.75.Dg, 37.25.+k}

\section{Introduction}

\begin{figure}
\begin{center}
\includegraphics[width=7cm]{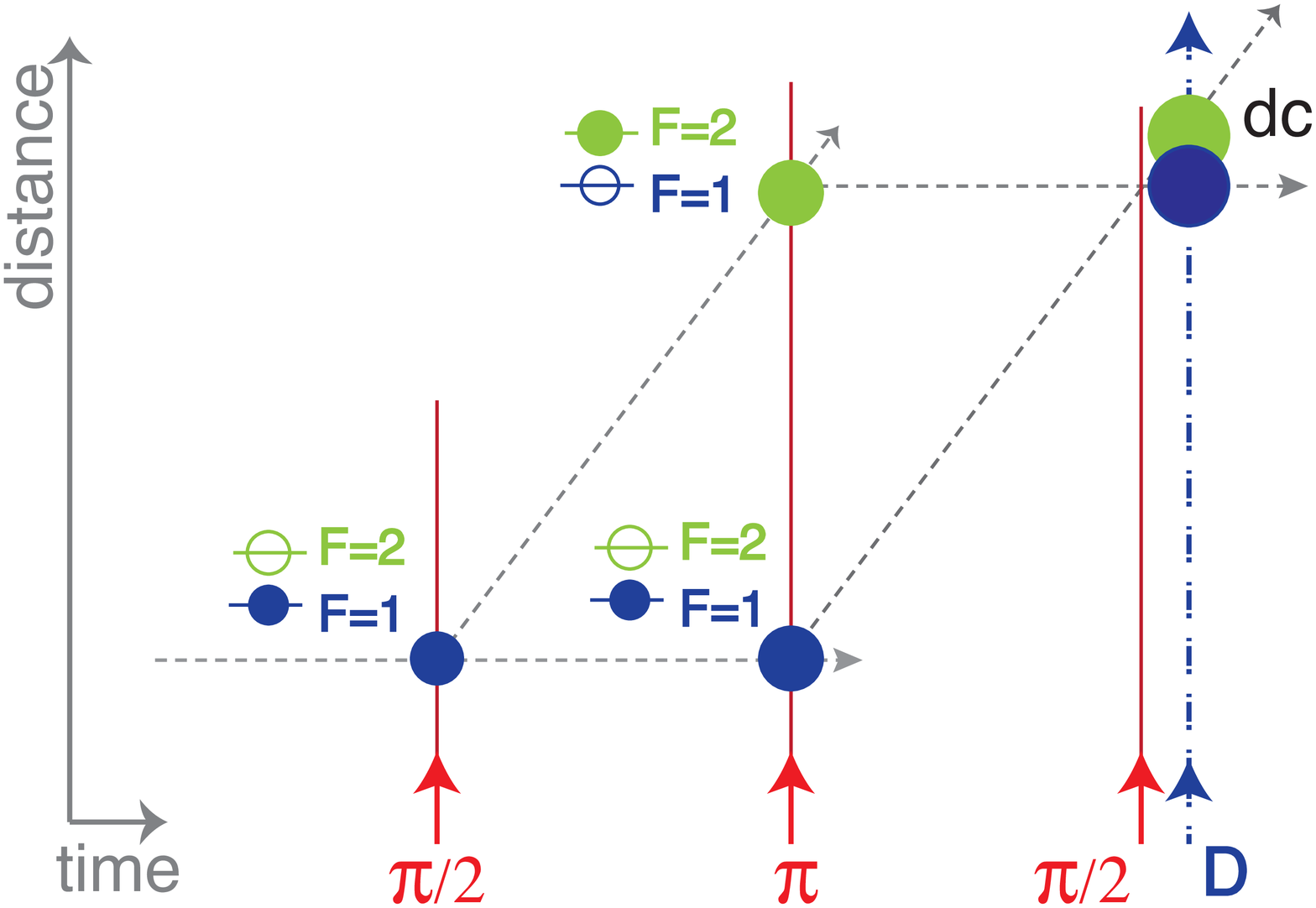}
\caption{Working schematics for a light pulse atom interferometer with $\pi$ and $\pi/2$ the beam pulses,  ``D'' the detection beam, and ``dc'' the output atom clouds.}
\label{AtomInterferometerScheme}
\end{center}
\end{figure}

In the late eighties  the invention of laser cooling \cite{Raab:1987va} pushed the development of atom interferometry (AI) \cite{Kasevich:1991vy}.  Since then  atom interferometers (AIs) have been used in basic science for the measurement of the $\hbar/m$ ratio  \cite{Weiss:1993wx}, of the fine structure constant \cite{Bouchendira:2011cc}, and the Newtonian gravitational constant \cite{Lamporesi:2008ut, Fixler:2007dj}, and for tests of  the equivalence principle of general relativity (EP) \cite{Fray:2004bt,Bonnin:2013gl}.  Fundamental science applications based on atom interferometry have been proposed such as, for example, very precise space EP tests \cite{Gaaloul:2010dh}, low frequency gravitational wave detectors \cite{Graham:2013ek} and laboratory tests of dark energy \cite{Perl:2010wm}. In applied science research and development efforts have been focused on the development of inertial and gravitational sensors based on AIs, such as accelerometers \cite{Geiger:2011hj}, gyroscopes \cite{Muller:2009fa}, gravimeters \cite{deAngelis:2011hi}, or gravity gradiometers \cite{Sorrentino:2013tf}. These quantum sensors will allow more precise measurements of Earth gravitational potential for climate-change studies, novel mapping of city undergrounds, and searches for new mineral and oil resources. \\
Atom interferometers rely on the optical control of the atom wave functions through the precise use and timing of laser pulses (see figure \ref{AtomInterferometerScheme}) and on the translation of the AI differential phase $\phi$ into the population unbalance among the possible energy, or momentum state of the output cloud atoms. In the case of figure \ref{AtomInterferometerScheme}, the expected renormalized population imbalance $\zeta$ is given by $(N_1-N_2)/(N_1+N_2)$, with $N_1$, and $N_2$ the atom number of the two AI output clouds, and it is a function of $\phi$. The population unbalance and therefore the AI differential phase are evaluated by  measuring the output cloud atom numbers during the final detection sequence.\\
The detection noise in the atom number measurement is one of the main noise sources limiting the sensitivity of AIs. Even if other noise sources might be present, as wavefront distortion, or phase noise in the AI laser pulses, being able to reduce the detection noise of the cloud atom number down to the atom shot noise limit is necessary to build AIs working at their optimal design sensitivity.\\
Some previous theoretical and experimental work has been carried out to examine the conditions to reach optimal detection for matter-wave optics \cite{Pappa:2011ig,Biedermann:2009tf}. However a detailed analysis of the errors affecting the detection noise of AIs, in particular with large atom number but small optical depth, is currently missing.
In section \ref{Section2} we study the technical challenges and error sources affecting fluorescence, absorption, and dark imaging detection for different detection sequences with the scope to identify a detection method which could be limited by the atom shot noise only.
In sections \ref{section3}-\ref{section4} we show how simultaneous fluorescence detection of both output clouds by CCD imaging poses less stringent technical requirements on the detection beam noise, and we identify the optimal detection parameters, like detection time, detuning, saturation ratio, CCD pixel size and number to reach the atom shot noise limit.
\section{Near resonant detection methods for atom interferometry}

\label{Section2}

\begin{figure}[htp]
  \centering
  \begin{tabular}{cc}
    \includegraphics[width=60mm]{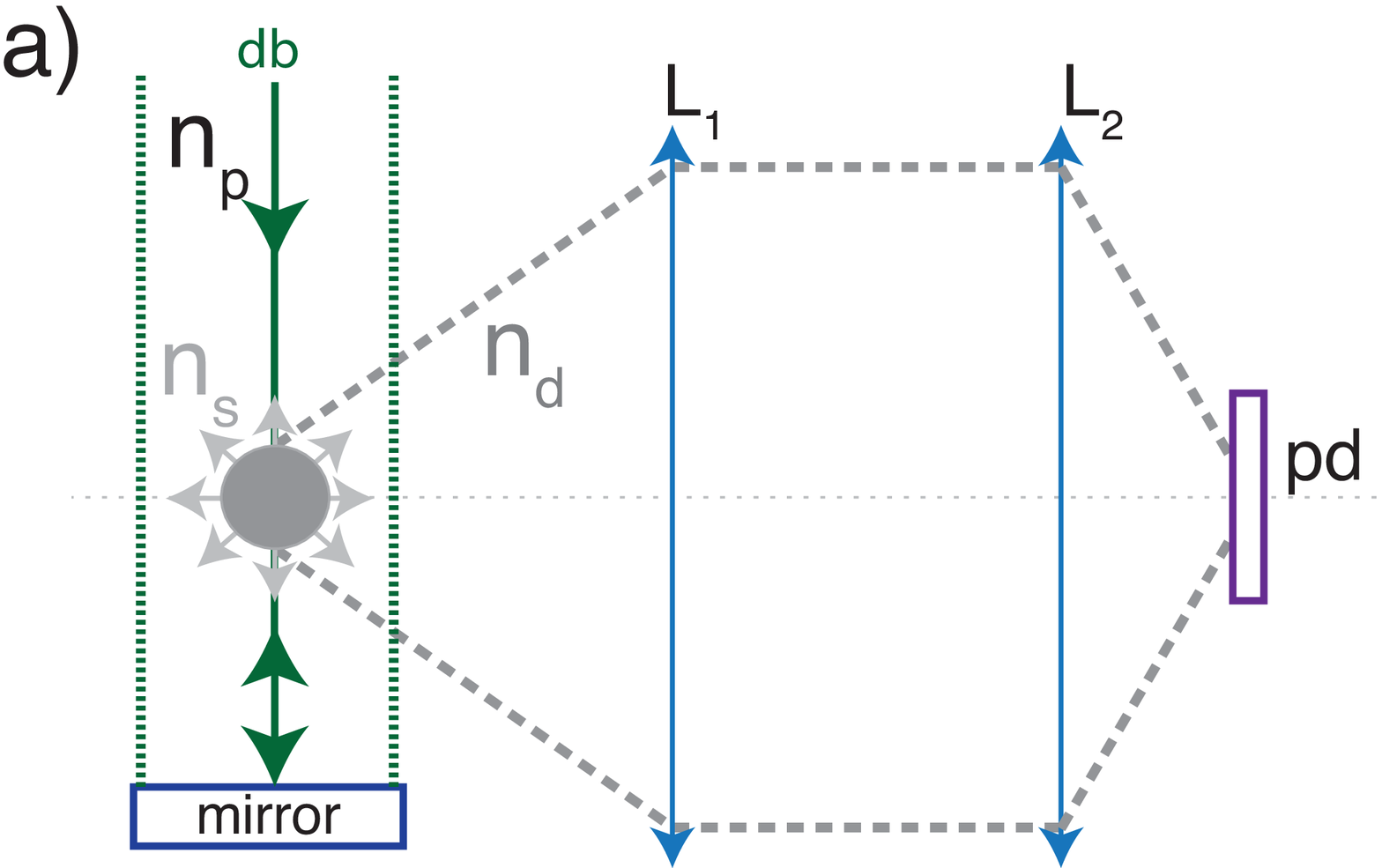}&
    \includegraphics[width=60mm]{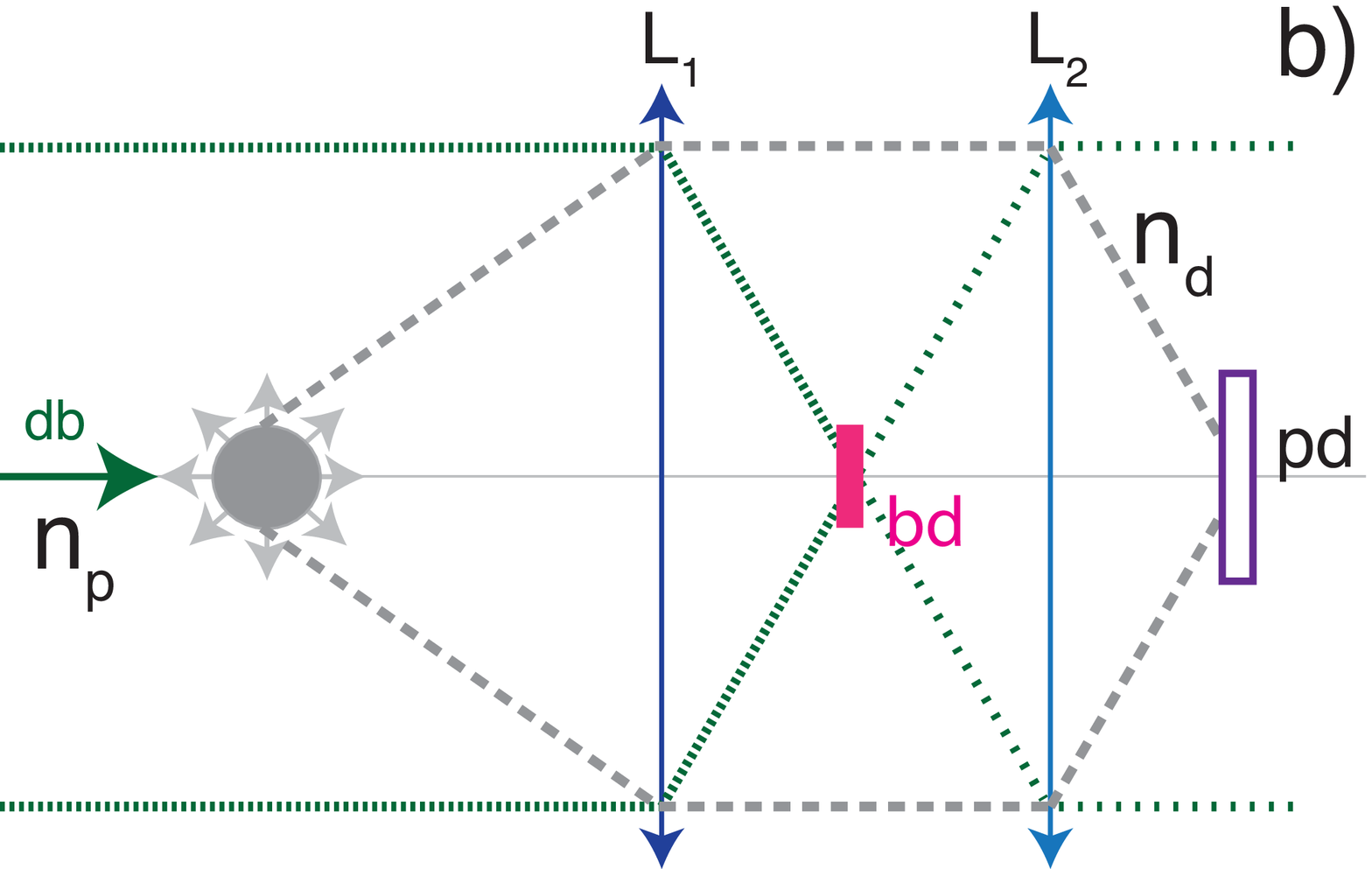}\\
  \end{tabular}
\caption{Detection scheme for : a) fluorescence; b) absorption and dark imaging, differing only by the presence of the blocking beam disk ``bd''. ``pd'' is the photodetector, ``db'' the detection beam, $L_1$, and $L_2$ a focusing lens system. $n_d$ is the detected photon number, $n_p$ the incoming photon number, and  $n_s$ is the scattered photon number.}
\label{DetectionMethods}
\end{figure}

Atom interferometers typically operate with large atom number, but optically thin clouds. To detect the atom number of these output clouds, the most often adopted detection methods in AI are near resonance ones, where a detection laser beam illuminates the atoms having a laser frequency, which is near the resonant frequency of the atomic transition to interrogate. The atom number is then detected by measuring the scattering dispersion, or absorption of photons, as in fluorescence, absorption, or dark imaging detection.  Other possible detection methods are far off resonance ones, which are based on the phase shift and polarization rotation induced by the atoms on the beam photons when these have a frequency far from the interrogated atomic resonance.  However, since these far off resonance methods are not often adopted in atom interferometry, we do not include them in this work.\\
In the near resonant methods the scattered and transmitted photons (see figure \ref{DetectionMethods}) are intercepted by a lens with numerical aperture NA, and then through some optics imaged onto a photo detector such as a CCD or a photodiode. Furthermore in dark-ground imaging, the detection beam is blocked by an additional disk positioned in the centre of the Fourier plane, and only the fluorescence and the diffracted light are imaged on the photodetector \cite{Pappa:2011ig}. From these measurements, once all the detection parameters are known, the cloud atom number $N$ can be derived.\\
This estimate of $N$ is affected by a number of noise sources, like atom shot noise and photon shot noise with standard deviations $\sigma_N=\sqrt{N}$ and $\sigma_p=\sqrt{n_d}$ respectively, where $n_d$ is the  photon number counted by the photodetector during the detection time.  The noises on laser intensity $I$ and frequency $\nu$ of the detection laser beam. We assume these noises to be Gaussian and with standard deviations $\sigma_I$ and $\sigma_{\Delta}$ respectively also induce an error on the estimate of $N$.\\
The noise optimization of the detection method depends very much on the atom number and the optical depth of the imaged clouds.  A typical AI has output clouds characterized by a small optical depth with $\sigma \rho_{2D}\ll1$ \cite{Pappa:2011ig}, where $\sigma$ is the atomic scattering cross section, and $\rho_{2D}$ the column atom density of the cloud along the beam axis. For a spherically symmetric Gaussian density distribution with standard deviation $\sigma_\rho$, the maximum column atom density along the detection beam optical axis is ${\rho_{2D}}_{max}=N/(2 \pi \sigma_\rho^2)$. In the following, we assume $N$ to be in the range $10^6-2\times 10^7$, and the cloud size $\sigma_\rho$ to be $5\textrm{mm}$. We also assume an imaging lens with a numerical aperture of $\textrm{NA}=0.2$, and to use $^{87}\rm{Rb}$ atoms. At resonance, for example, for circularly polarized light, $\sigma$ is $\sigma_0=2.9\times 10^{-13}m^{-2}$ \cite{Steck:Rubidium87Data}. Under these assumptions because of $\sigma_0 {\rho_{2D}}_{max}\approx2\times10^{-3}$ the hypothesis of small optical depth is valid.\\
Given the assumed small optical depths, we can exclude the dark-ground imaging from the analysis. In fact, for $\sigma \rho_{2D}\ll1$ the fluorescence signal dominates the coherent diffraction at near-resonance \cite{Pappa:2011ig}, and dark ground imaging is fully equivalent to fluorescence detection.\\
Absorption imaging is also excluded because of the technical challenges to image enough photons on a CCD and reach the atom shot noise limit.
Indeed, only for a minimum number of scattered detected photons $(n_d)_{min}$,  the contribution to the detection noise of the atom shot noise $\sigma_N$ dominates over the contribution of the photon shot noise $\sigma_{n_d}$ by a factor  $\beta$ (we assume $\beta=3$). For low optical depths, we are now showing that  $(n_d)_{min}$ for absorption is much larger than $(n_d)_{min}$ for fluorescence, and technically difficult to detect by CCD imaging.
In absorption imaging one usually measures the transmittance $T$ as defined by $n_d/n_p$ with $n_p$ the number of the incoming photons, and $n_d$ the photons number reaching the detector, which is given by the equation \cite{Pappa:2011ig}:
  \begin{equation}
{n_d} = {n_p}\exp ( - \sigma {\rho _{2D}}) \approx {n_p}(1 - kN)
\end{equation}
with $k=\sigma {\rho _{2D}}/N$, where $\sigma$ is the scattering cross section (see equation (\ref{sigmaDel})). The total noise on the measured transmittance $T$ due to the atom and photon shot noises is equal to $\sigma_T=\sqrt{2 \sigma_p+(n_p\kappa)^2 \sigma_N^2 }$.  Considering also that the transmittance $T$ for a cloud with a Gaussian density profile has a minimum value $T_{min}\approx1- k_{max} N$ with $k_{max}=\sigma /(2\pi \sigma_\rho^2)$, the atom shot noise contribution to $\sigma_T$ is dominant over the contribution of the photon shot noise by at least a factor  $\beta$ only for $n_d>(n_d)_{min}=2\beta^2/{( k_{max}^2 N)}$. 
Under our numerical assumptions, and with $N=10^6$,  $(n_d)_{min}$ is $\approx5.2\times10^{11}$, which, assuming a CCD sensor with a maximum pixel well depth of $10^5 \textrm{photons/pixel}$, requires a minimum pixel number of $5.2 \times 10^6$ per cloud to be imaged. For an AI with two or more clouds (as in \cite{Gaaloul:2010dh}) to be imaged, CCDs with such a pixel numbers are not yet easily available on the market.\\
On the other hand $(n_d)_{min}$ for fluorescence is easily detectable with a CCD. Indeed, for small optical depths \cite{Pappa:2011ig}, we have that $n_d=\alpha N$ with $\alpha$ a proportionality constant depending on the atom cross section and the duration and intensity of the probe beam (see equation (\ref{ndtwolevelFull})). The total detection noise on $n_d$ for fluorescence is $\sqrt{\sigma_p^2+\alpha^2 \sigma_N^2 }$, which,  independently of the expression of $\alpha$, has the atom shot noise contribution dominant by a factor $\beta$ over the photon shot noise for $n_d>(n_d)_{min}=N \beta^2$.
It is interesting to notice that for fluorescence detection $(n_d)_{min}$ increases with the atom number, which is the opposite to what is required for absorption detection.
With our assumptions and for fluorescence detection $(n_d)_{min}$ is $ 9\times 10^6$, which requires only a $90$ pixel CCD assuming a pixel well depth of $10^5 \textrm{photons/pixel}$. \\
Even if detection with a photodiode is technically simpler than with a CCD, and allows shot noise limited absorption detection, using a CCD is still preferable since it allows one to record the spatial information of the output clouds. This information can be useful for some applications, because it can be used to estimate the angular velocity of the AI \cite{Sugarbaker:2013bk,Dickerson:2013ur}. It also allows one to resolve different output clouds, if they are simultaneously imaged.
For atom interferometry, and CCD imaging, fluorescence is the less challenging, and therefore our preferrred detection method to measure the output cloud numbers at the atom shot noise limit. 
\section{Sequential versus simultaneus fluorescence detection }

\label{section3}

Given fluorescence as the chosen detection method, the AI output clouds can be imaged in a sequential or simultaneous way: in sequential imaging, the hyperfine state populations of the atoms in the AI output clouds are estimated in a time sequence one after the other one; in simultaneous imaging the atoms in different hyperfine states are in spatially distinguishable clouds which are imaged at the same time. Sequential imaging is simpler to realize than the simultaneous one, but, as we discuss in section \ref{Section:ConstraintsDetTimeSequential}, because of atom loss to unwanted transitions it requires a longer interrogation time, and is constrained by a maximum allowed laser detuning with respect to the resonance. As we discuss in section \ref{Section:ConstraintsLaserNoiseSequential}, it is also affected by the laser frequency and intensity noise, requiring dedicated detection beam stabilization.  As we show in section \ref{Section:AdvantagesAndConstrainsSimImag}, simultaneous fluorescence detection with CCD imaging instead allows one to measure the atom cloud population at the atom shot noise limit with less stringent technical requirements on the stability of the detection-beam parameters.\\
\subsection{\bf Limitations of sequential Imaging}
In sequential imaging, the AI output clouds are imaged in two ways: if the clouds are spatially separated, they are imaged one after the other one as they fall under gravity illuminated by one detection beam in front a photo detector; otherwise the superimposed clouds have the atoms in distinct hyperfine ground states, and their populations are imaged one after the other by using different detection beams tuned to different detection frequencies. 
\subsubsection{\bf Constraints on detection time, and detuning frequency}

\label{Section:ConstraintsDetTimeSequential}
\begin{figure}
\begin{center}
\includegraphics[width=8cm]{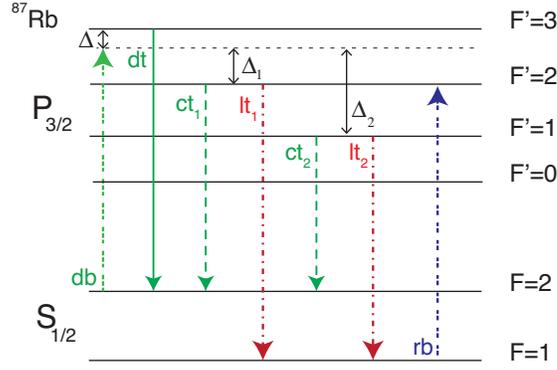}
\caption{Schema of the modelled transitions for the $^{87}\textrm{Rb}$ $D_2$ lines. $lt_1$, and $lt_2$ are the loss transitions, ``dt'' the detection one, $ct_1$, and $ct_2$ are closed depumping transitions. ``db'' is the detection beam, and ``rb'' the repumper one.}
\label{TransitionsLosses}
\end{center}
\end{figure}

\begin{figure}
\begin{center}
\includegraphics[width=7.5cm]{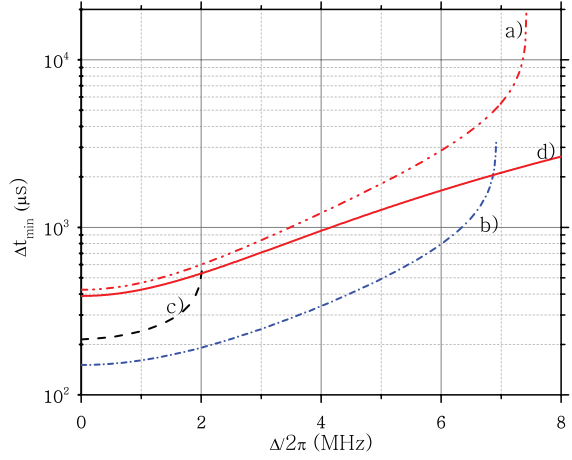}
\caption{ Minimum detection time $(\Delta t)_{min} $ for $^{87}\textrm{Rb}$ for different saturation ratio $s$: ``a'', ``b'', and ``c'' refer to $s$ of 0.2, 1, 6.  The continuous line ``d'' is  ${\left[ {{{(\Delta t)}_{min}}} \right]_{2l}} $ for the two atom level model for $s=0.2$}
\label{DtMin}
\end{center}
\end{figure}

\begin{figure}
\begin{center}
\includegraphics[width=7.3cm]{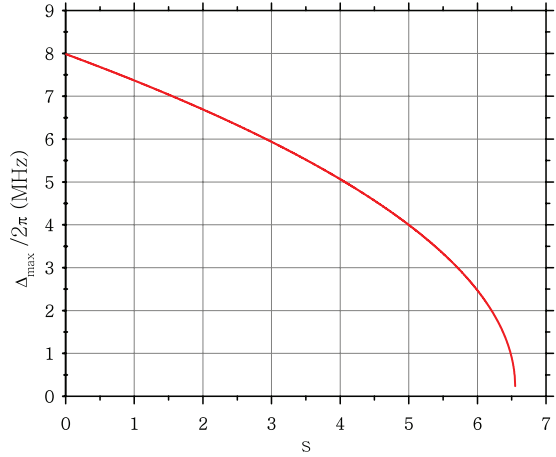}
\caption{Maximum detuning as a function of the saturation ratio $s$ for $^{87}\textrm{Rb}$.}
\label{MaxDet}
\end{center}
\end{figure}

In the two level model of detection (see \ref{AppendixnsTwoLevel}), for the example of $^{87}\textrm{Rb}$ the atoms in a hyperfine $F=2$ ground state cycle through the excited $F'=3$ state emitting the photons to be detected by the photodetector.  However, in a realistic  multi-level model, some atoms might be excited to secondary hyperfine states, and then subsequently lost through non closed transitions. In the case of $^{87}\textrm{Rb}$  atoms with the $D_2$ line hyperfine structure represented in figure \ref{TransitionsLosses} \cite{Steck:Rubidium87Data}, a detection beam with frequency detuned by $\Delta$ with respect to the $F=2 \rightarrow F'=3$ transition induces not only the desired fluorescence transition, but also the $F=2 \rightarrow F'=2$, and $F=2 \rightarrow F'=1$ ones. As represented in figure \ref{TransitionsLosses} (dashed red arrow) some of the atoms in these $F'\neq 3$ excited states spontaneously decay to the $F=1$ ground state, which is decoupled from the detection light. If the AI output state clouds are physically superimposed, a repumper beam can not be used to cycle back the atoms from the $F=1$ to the $F=2$ ground state. In this case the leakage to the $F=1$ state is effectively an atom loss for the AI population estimation, and modifies the number of scattered photons versus time, $n_s(t)$ of the two level atom. By modifying $n_s(t)$, this atom loss implies an increase of the minimum detection time $(\Delta t)_{min}$, and puts constraints on the largest detuning $\Delta$ allowed, and the saturation ratio $s=I/I_{sat}$ to reach the atom shot noise limit, where $I_{sat}$ is the saturation intensity  as defined  \cite{Steck:Rubidium87Data,Steck:moWuU-GK}.
In the model which includes atom loss through non closed transitions, $(\Delta t)_{min} $ is given by the following equation:
 \begin{equation}
{(\Delta t)_{min}} = \frac{{{I_{sat}}}}{I}\frac{2}{{\Gamma {\gamma _{lost}}}}\log \left[ {{{\left( {1 - \frac{{{\gamma _{lost}}}}{{{\gamma _d}}}\frac{{{{({n_s})}_{min}}}}{N}} \right)}^{ - 1}}} \right],
\label{DtMinWithLosses}
\end{equation}
 where $\Gamma$ is the natural line-width (FWHM) of the considered transitions, which, in the case of $^{87}\textrm{Rb}$, is $2\pi\times6.06 \textrm{MHz}$ \cite{Steck:Rubidium87Data}. The parameters $\gamma_d=\gamma(\Delta)$ and $\gamma_{lost}=\sum_i \gamma (\Delta i) \textrm{br}_i$ are respectively proportional to the scattering cross sections of the detection fluorescence transition and of all  the lost non closed transitions. The function $\gamma(\Delta)$  is the function given by:
 \begin{equation}
\gamma (\Delta ) = {\left( {1 + \frac{I}{{{I_{sat}}}} + 4\frac{{{\Delta ^2}}}{{{\Gamma ^2}}}} \right)^{ - 1}},
\end{equation}
where $\Delta_i=\Delta_{di}-\Delta$ is the detuning with respect to the non detection hyperfine transitions, and $\Delta_{di}$ is the frequency difference of these transitions with respect to the detection one. Finally $\textrm{br}_i$ is the spontaneous emission branching ratio to not closed transitions (see \ref{AppendixBranchingRatio}). From section \ref{Section2}, $(n_s)_{min}$ is given by $(n_d)_{min}/\Omega$ with $(n_d)_{min}=N \beta^2$, where $\Omega \approx \textrm{NA}^2/4$ is the fraction of solid angle intercepted by the imaging lens of numerical aperture NA. \\
In some cases the requirements on the photon shot noise are too strict, and $(\Delta t)_{min}$  does not exist because all the atoms are lost to not closed transitions before scattering enough photons to reach the atom shot noise limit.  Given a choice of $\beta$ and saturation ratio $s$ this happens for detuning larger than some detuning $\Delta_{max}$, which is found by numerically solving the following equation in $\Delta_{max}$:
 \begin{equation}
\frac{{{\beta ^2}}}{{\Omega}}= \frac{{{\gamma _d}({\Delta _{max}})}}{{{\gamma _{lost}}({\Delta _{max}})}}.
\end{equation}
Equivalently for a chosen $\beta$ and detuning parameter $\Delta$, there is a maximum allowed saturation ratio $s_{max}$ to reach the atom shot noise limit.\\
For the alkali atoms, small saturation ratio $s$ and small detuning $\Delta$, $\gamma_{lost}/\gamma_d$ is of the order of $10^{-3}$.  Under these conditions $(\Delta t)_{min} $  can be described by the equation:
 \begin{equation}
{(\Delta t)_{min}} \approx {\left[ {{{(\Delta t)}_{min}}} \right]_{2l}}\left[ {1 + \frac{{{\gamma _{lost}}}}{{{\gamma _d}}}\frac{{{\beta ^2}}}{{\Omega}}} \right]
\label{DtMinWithLosses2}
\end{equation}
with ${\left[ {{{(\Delta t)}_{min}}} \right]_{2l}}$ the minimum detection time to reach the atom shot noise limit for the two level model with atom loss (see \ref{AppendixnsTwoLevel}) given by the equation:
 \begin{equation}
{\left[ {{{(\Delta t)}_{min}}} \right]_{2l}} = \frac{2}{\Gamma }\frac{{{I_{sat}}}}{I}\frac{1}{{{\gamma _d}}}\frac{{{\beta ^2}}}{{\Omega}}.
\label{DtminSimultaneus}
\end{equation}
For example, assuming $\textrm{NA}=0.2$ and $\beta=3$, equation (\ref{DtMinWithLosses2}) implies an increase of the minimum detection time due to the atom loss of at least a factor two. This increase of $(\Delta t)_{min} $ progressively gets larger for larger detuning and saturation ratio, because more atoms are  lost to non closed transitions. 
Under our numerical assumptions, in figure \ref{DtMin} we plot $(\Delta t)_{min}$ versus the detuning $\Delta$ with respect to the $F'=3$ hyperfine level of $^{87}\rm{Rb}$ for different saturation ratios $s$. The branching ratio to the not closed transition is $50\%$ and $83\%$ for respectively the hyperfine level $F'=2$ and $F'=1$ (see table \ref{BranchingRatioTable}).  From figure \ref{DtMin}, it is clear that for detuning $\Delta \geq 2 \pi \times 7.4 \textrm{MHz}$, it is impossible to reach atom shot noise detection. For comparison, we plot as well the curve for the simple two level system of equation (\ref{DtminSimultaneus}) as a continuous line, which shows a much reduced detection time for large detuning. In figure \ref{MaxDet} we plot the maximum allowed detuning $\Delta_{max}$ as a function of the saturation ratio $s$ showing that, for large detunings, it is necessary to have low intensity detection beam to reach the atom shot noise limit. The same figure  means that, as a function of the detuning, there is a maximum allowed saturation ratio.  
Clearly a detection sequence, like simultaneous imaging with a repumper beam, which has the same detection time of the simple two level model, is favorable with respect to the one of the sequential imaging.  

\subsubsection{\bf Constraints on frequency, and power noise of detection beam}

\label{Section:ConstraintsLaserNoiseSequential}
\begin{figure}
\begin{center}
\includegraphics[width=7.5cm]{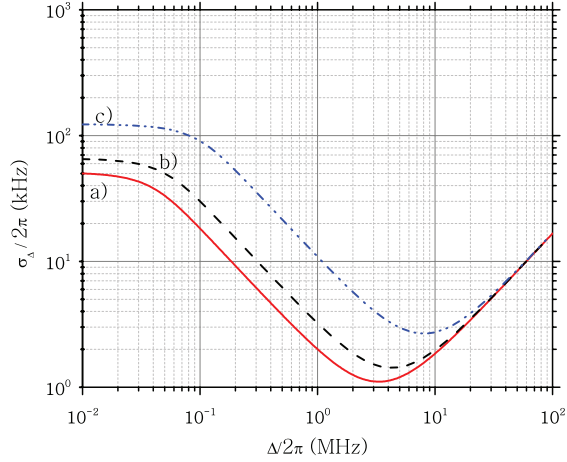}
\caption{Requirements on the laser frequency noise $\sigma_\Delta$ versus the detuning frequency $\Delta$ for $T=1(\textrm{s})$ for different value of saturation ratio: ``a'', ``b'', and ``c'' refer to values of $s$ of 0.2, 1, and 6 respectively. }
\label{LaserFrequencyNoise}
\end{center}
\end{figure}

The laser frequency and power noises also induce noise on the detection. We show now that for sequential detection the requirements on these noise sources to reach the atom shot noise limit can be technically challenging, and make simultaneous detection a preferable alternative detection method (see section (\ref{Section:AdvantagesAndConstrainsSimImag})). 
If $\sigma_\Delta$ is the standard deviation of the laser frequency noise between the readouts of the output clouds, the requirements on $\sigma_\Delta$ for small detuning $\Delta$ to reach the atom shot noise limit are given by:
 \begin{equation}
{\sigma _\Delta } < \Gamma \frac{{\sqrt {1 + \frac{I}{{{I_{sat}}}}} }}{{\sqrt \beta  2{{\left( {2N} \right)}^{1/4}}}}
\label{eqfreqreqnoiselowDelta}
\end{equation}
and for larger $\Delta$ by:
 \begin{equation}
{\sigma _\Delta } < \frac{1}{{\beta \sqrt N }}\frac{{{\Gamma ^2}}}{{8\Delta }}\left( {1 + 4\frac{{{\Delta ^2}}}{{{\Gamma ^2}}} + \frac{I}{{{I_{sat}}}}} \right).
\label{eqfreqreqnoisehighDelta}
\end{equation} 
Similarly the requirements on the fractional standard deviation $\sigma _I/I$ of the laser intensity noise between the output readings to reach the atom shot noise limit are given by the equation:
 \begin{equation}
\frac{{{\sigma _I}}}{I} < \frac{1}{{\beta \sqrt N }}\left( {1 + \frac{I}{{{I_{sat}}}}\frac{1}{{1 + 4\frac{{{\Delta ^2}}}{{{\Gamma ^2}}}}}} \right).
\label{eqIntensityNoise}
\end{equation}
The previous constraints on the laser noises have been derived from their induced error on the detected photons, $n_d$. For the  laser frequency noise $\sigma_\Delta$, its induced error on $n_d$, ${(\sigma _{{n_d}}^2)_{{\sigma _\Delta }}}$, is given by the equation:
 \begin{equation}
{(\sigma _{{n_d}}^2)_{{\sigma _\Delta }}} = {\left( {{{\partial \alpha } \over {\partial \Delta }}N} \right)^2}\sigma _\Delta ^2 + {\left( {{{{\partial ^2}\alpha } \over {\partial {\Delta ^2}}}N} \right)^2}{{\sigma _\Delta ^4} \over 2}
\label{eqallfrequencynoises}
\end{equation}
with the second order derivative of  $\alpha$ with respect to $\Delta$ \cite{Papoulis:2002wq} being the dominant term for low detuning. By requiring that $\alpha \sigma_N>\beta(\sigma_{n_d})_{\sigma_\Delta}$, the requirements on $\sigma_\Delta$ of equations (\ref{eqfreqreqnoiselowDelta}-\ref{eqfreqreqnoisehighDelta}) can be derived. 
With a similar approach  the requirements on the laser intensity noise of equation (\ref{eqIntensityNoise}) can be derived as well. 
In figure \ref{LaserFrequencyNoise} we plot the maximum laser frequency noise $\sigma_\Delta$ for $^{87} \rm{Rb}$ with the numerical assumptions of section \ref{Section2}.  In this figure, the curve on the requirements on $\sigma_\Delta$  is mainly given by the first derivative of $\alpha$ in $\Delta$ which has a local maximum in correspondence to the inflection point of  $\alpha$ as a function of $\Delta$.
Considering our numerical assumptions,  to reach the atom shot noise limit $|\sigma_I/I|$ has to be smaller than $0.04\%$ for zero detuning and $0.033\%$ for large detuning. \\
 \begin{figure}
\begin{center}
\includegraphics[width=7.5cm]{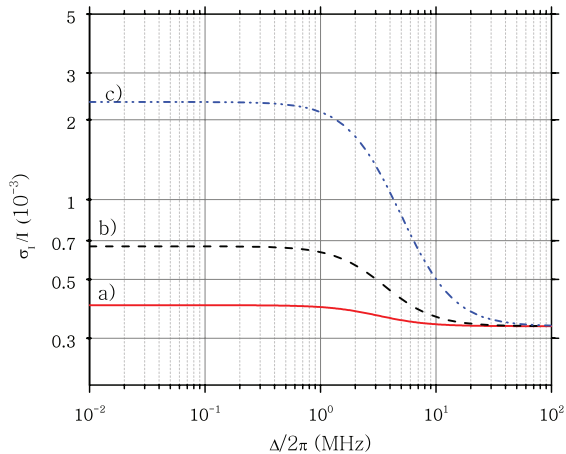}
\caption{Requirements on the fractional standard deviation $\sigma_I / I$  for the different values of the saturation ratio: ``a'', ``b'', and ``c'' refer to the value of $s$ of 0.2, 1, and 6 respectively. }
\label{ImageASDIntensityNoise}
\end{center}
\end{figure}
Even if a long detection time works effectively as a low pass filter, which averages out high frequency noise of the detection beam, the low frequency components still affect the detection.  
These requirements on laser frequency and intensity noise need electronic stabilization and it has already been done in some experiments \cite{Sorrentino:2013tf}. This electronic stabilization however can prove to be technically challenging because the relative long time between the detection pulses requires stabilization at low frequencies, where technical noise tends to dominate.  \\
Therefore alternative approaches should be explored to reduce these laser noise requirements, and to allow short detection times. 
\subsection{\bf Advantages and constraints of simultaneus imaging}

 \label{Section:AdvantagesAndConstrainsSimImag}
An alternative approach to detect the output clouds of the atom interferometer is to image them all during the same detection time. During this simultaneous imaging, the output clouds have to be separated in space to be resolved, otherwise before detection a pusher beam is used to physically separate them. During detection, a repumper beam is used too to transfer all the atoms of the output clouds to the same hyperfine ground level as shown in figure \ref{TransitionsLosses}, so that a common detection beam can be used to detect both clouds. The use of  this repumper beam prevents any atom loss through non closed transitions as it happens for sequential imaging, and decreases the minimum detection time $(\Delta t)_{min}$ required to reach the atom shot noise limit. Simultaneous imaging introduces a new noise source associated with cross cloud shadowing, which however we show in section \ref{CrossShadowingSection} still allows one to reach the atom shot noise limit.
Simultaneous imaging does not require as much laser stabilization as sequential imaging, and results in a less constrained detection scheme. Because the AI phase is a function of the population unbalance between the AI outputs, $\zeta$, and so it is a relative measurement, the additional complexity of using a repumper beam comes with the advantage that simultaneous imaging allows common mode rejection of the effects of the laser frequency and intensity noises of the detection and repumper beams in the AI phase estimate.\\

\subsubsection{\bf Reduced constraints on detection time}
\begin{figure}
\begin{center}
\includegraphics[width=7.5cm]{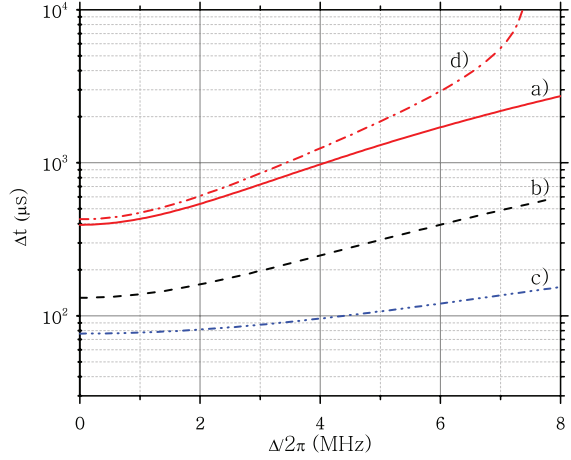}
\caption{Minimum detection time ${\left[ {{{(\Delta t)}_{min}}} \right]_{2l}} $ for different saturation ratio $s$: ``a'', ``b'', and ``c'' refer to the values of $s$ of 0.2, 1, and 6 respectively.  The continuous line ``d'' is  $(\Delta t)_{min} $ for the two level atom model with losses for $s=0.2$}
\label{DtMinSimultaneus}
\end{center}
\end{figure}

Simultaneous imaging thanks to the repumper beam can be modeled with the simple two level atom (see \ref{AppendixnsTwoLevel}), and has its minimum detection time ${\left[ {{{(\Delta t)}_{min}}} \right]_{2l}}$ given by equation (\ref{DtminSimultaneus}), which does not depend on the number of atoms to detect.
For example, with the assumptions of section \ref{Section2}, and a saturation ratio $s=0.2$,  $\left[ {{{(\Delta t)}_{min}}} \right]_{2l} $ is $\approx 400 \mu \textrm{s}$ for zero detuning, and, as shown in figure \ref{DtMinSimultaneus}, increases with the detuning, but becomes smaller for larger saturation ratios.\\
\subsubsection{\bf No constraints from cross cloud shadowing}

\label{CrossShadowingSection}
\begin{figure}
\begin{center}
\includegraphics[width=8cm]{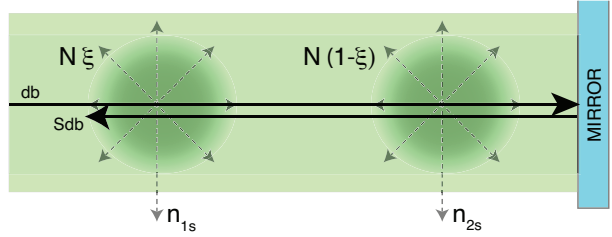}
\caption{ Simultaneous readout of two clouds. ``bd'' is the detection beam, ``Sdb'' the shadowed detection beam, $n_{1s}$, and $n_{2s}$ the scattered photon number.} 
\label{TwoCloudShadows}
\end{center}
\end{figure}

In simultaneous imaging, the output clouds share the same detection beam, and, as shown in figure \ref{TwoCloudShadows}, cast shadows onto each other. These shadows effectively reduce the number of scattered photons with respect to the shadow-free case, introducing a systematic error.
The systematic error on the estimate of the population unbalance between the two AI output cloud $\zeta$ induced by the cross cloud shadowing, $\delta_\zeta$, is given by the difference between the ``real’’  value $\zeta$, and the measured one. This measured population unbalance is given by $\zeta_m=(n_{1s}-n_{2s})/(n_{1s}+n_{2s})$ with $n_{1s}$ and $n_{2s}$ the photons scattered from the first and second output cloud. This approach to estimate $\zeta$ cancels out the effect of the atom number fluctuations between experimental runs, hence reducing the overall phase noise. For small optical depths, $\delta _\zeta$ is a function of the relative atom number of one output cloud, $\xi=N_1/N$, with respect to the total atom number $N=N_1+N_2$, and  is given by (see \ref{AppendixShadow}):
 \begin{equation}
{\delta _\zeta } =  - 2\Lambda \frac{{\xi (1 - \xi )(1 - 2\xi )}}{{1 - \tau  + \Lambda (1 - \xi (1 - \xi ))}},
\label{eq:errorFromShadowingNormalized}
\end{equation}
where $\tau$ is twice the optical depth associated with the maximum column density ${\rho _{2D}}_{max}$ and $\Lambda$ is defined in \ref{AppendixShadow}.
In figure \ref{CrossCloudsShadows}, for the  assumptions of section \ref{Section2}, with $N=3\times10^6$, $\Delta=0$, and $s=0.2$, we plot the error on the population unbalance ${\delta _\zeta }$ in unit of the error on $\zeta$ due to the atom shot noise only, $(\sigma_{\zeta})_{\sqrt{N}}$, versus the values of $\xi$. The error $(\sigma_{\zeta})_{\sqrt{N}}$ is itself a function of $\xi$, and equal to $2\sqrt{\xi(1-\xi)/N}$. The plotted function ${\delta _\zeta }/(\sigma_{\zeta})_{\sqrt{N}}$ has a local maximum at $\xi\approx0.15$, which shows that the cross-cloud shadowing induced error is just $\approx4\%$ of the error coming from the atom shot noise. Cross cloud shadowing therefore does not substantially limit the detection. \\ 
 \begin{figure}
\begin{center}
\includegraphics[width=7.5cm]{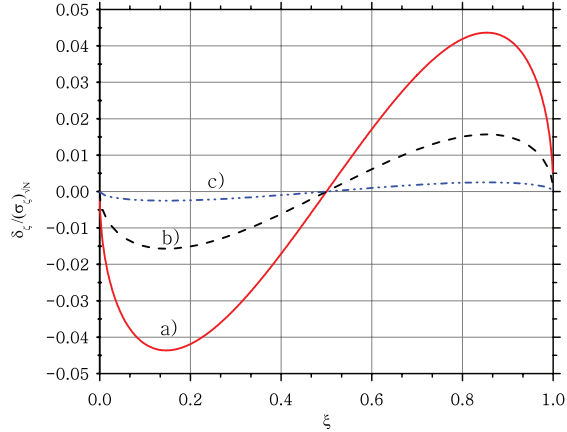}
\caption{ Error on the population unbalance ${\delta _\zeta }$ in unit of the error on $\zeta$ due to the atom shot noise, $(\sigma_{\zeta})_{\sqrt{N}}$, as a function of $\xi$, and of the saturation ratio $s$ on resonance: ``a'', ``b'', and ``c'' refer to the values of $s$ of 0.2, 1, and 4 respectively.}
\label{CrossCloudsShadows}
\end{center}
\end{figure}
Simultaneous fluorescence imaging thus puts less stringent constraints on the detection beam noises, and needs a smaller detection time than sequential imaging to reach the atom shot noise limit. \\
\section{CCD as imaging sensor for simultaneus imaging}

\label{section4}

The previous analysis assumed the choice of CCD imaging over photodiode imaging. Indeed CCD imaging allows one to record the spatial information of the output clouds. As shown below, this spatial information relaxes the requirements on the minimum distance between the output clouds to avoid cross readouts down to the atom shot noise limit. Furthermore, as shown recently \cite{Sugarbaker:2013bk,Dickerson:2013ur},  the recorded spatial information can be used to characterize velocity dependent phase shifts in single shot measurements,  either to improve AI sensitivity or to design new measurement schemes. 
\subsection{\bf CCD to record the cloud spatial information}
In simultaneous imaging, the output clouds, even if they are physically separated, have the tails of their distribution partially overlapped. If the spatial information of the clouds is not preserved, but only the volume integrals of the photon number is recorded (e.g. with a single photodiode), this cloud overlap induces a systematic error on the population measurements. This error is smaller than the atom shot noise contribution by a factor $\beta$ only if the output cloud centers are at a distance larger than ${d_{min}(\beta)}$, which can be large,  and technically difficult to achieve. The spatial information recorded by a CCD can instead be used to spatially fit the atom cloud distribution, and so estimate their population without introducing the previous systematic error. 
For two clouds imaged in two semi planes by a photodiode or photo multiplier, which have Gaussian density distributions with equal $\sigma_\rho$ and equal total number $N/2$, ${d_{min}}$  is given by:
 \begin{equation}
{d_{min}} =2\sqrt 2 {\sigma _\rho }{\rm{ erf}}^{ - 1}\left( {\frac{{2\sqrt 2 }}{{\beta \sqrt N }}} \right),
\label{eq:dminoverlap}
\end{equation}
where $\rm{erf}^{-1}$ is the inverse of the error function. For our numerical assumptions, the previous equation can be approximated with:
 \begin{equation}
d_{min}\approx2 \sigma_\rho \sqrt{\zeta-\ln \left(\zeta\right)}
\end{equation}
with 
 \begin{equation}
\zeta  = \ln \left( {2/\pi } \right) - 2\ln \left( {\frac{{2\sqrt 2 }}{{\beta \sqrt N }}} \right).
\end{equation}
For $N=10^{6}$ and our assumptions on ${\sigma _\rho }$ and $\beta$, ${d_{min}}$ is equal  to $6.6\sigma_\rho$, or $3 \textrm{cm}$, which for certain applications can be too restrictive. \\
Therefore, for simultaneous cloud detection, CCD imaging, because both it is less technically demanding and allows one to extract further AI information, is preferable over photodiode imaging.
\subsection{\bf CCD requirements}
Now given some CCD parameters such as the pixel well depth $n_{w} $, the pixel readout noise $\sigma_{px}$, the standard deviation of the dark signal per second $\varsigma _{dk}$, pixel binning number $m$, and the quantum efficiency $\mu$, we identify the minimum and maximum CCD pixel numbers to achieve the atom shot noise limit.   In fact a CCD needs to have at least $  (N_{\textrm{CCD}})_{min}$ pixels to record the fluorescence signal $(n_d)_{min}$:
 \begin{equation}
{({N_{\textrm{CCD}}})_{min}} = \frac{{N{\beta ^2}}}{{{n_w}m}}.
 \end{equation}
The maximum pixel number ${({N_{\textrm{CCD}}})_{max}}$ for a detection time $\Delta t$ is instead given by the equation:
 \begin{equation}
{({N_{\textrm{CCD}}})_{max}} = N{\beta ^2}{\mu ^2}{\left( {\frac{{\sigma _{px}^2}}{m} + \varsigma _{dk}^2\Delta {t^2}} \right)^{ - 1}}.
\end{equation}
This expression for ${({N_{\textrm{CCD}}})_{max}}$ has been obtained by requiring that the total readout noise, ${\sigma _{\textrm{CCD}}}^2 = {N_{\textrm{CCD}}}{\left( {{\sigma _{\textrm{CCD}}}/\mu } \right)^2}$, is equal to the photon shot noise associated with $(n_d)_{min}$. 
For example, if we assume $n_{w}=10^4 \textrm{photons/pixel}$, $\mu=0.7$,  $\sigma_{px}\approx 5 \textrm{electrons}$, $\varsigma _{dk}=10^3 \textrm{electrons/s}$, $\Delta t=500 \mu \textrm{s}$, $m=1$ and $N=10^6$ we find that $({({N_{\textrm{CCD}}})_{min}},{({N_{\textrm{CCD}}})_{max}})\approx (900,174\times10^3)$. For these pixel numbers and, for example, a CCD size of $13.3\textrm{mm}\times13.3\textrm{mm}$, the pixel size should be between $31\mu \textrm{m}$ and $443 \mu \textrm{m}$. 
These technical requirements on the CCD to reach the atom shot noise limit are achievable in commercial scientific sensors.\\

\section{Conclusions}

In this work we identified fluorescence simultaneous detection of  the AI output clouds by CCD imaging as the optimal detection method which poses the least stringent requirements on the detection beam noise and duration. This detection method also allows one to record the cloud spatial information, key for improving  future AI designs and sensitivity \cite{Sugarbaker:2013bk,Dickerson:2013ur}, and to reach the atom shot noise limit.  
 AI detection at the atom shot noise limit can allow the further characterization and reduction of other non detection noise sources limiting the AI sensitivity, like for example wavefront aberration, and Coriolis phase shifts.
The atom shot noise limit can then be lowered by using much larger atom numbers than the ones considered in this work, but further detection optimization will be required given the increased optical densities. Otherwise the AI sensitivity can be enhanced by other techniques, such as, for example, by using large area interferometers with large momentum transfer beam splitters \cite{Chiow:2011eu}, or by using squeezed atomic states \cite{Eckert:2006gq}, and reach sensitivities below the classical atom shot noise limit. 
Atom interferometry with high sensitivity at or below the atom shot noise limit will so open new science frontiers such as, for example, tests of general relativity equivalence principle on quantum matter \cite{Aguilera:2013wy}. Furthermore, high precision AIs will be the next generation inertial and gravitational sensors used  for civil engineering, geophysical exploration, or climate studies, opening up a new range of possible application. 
\ack
We thanks EPSRC for support of the GGTop project, grant number EP/I036877/1, the Royal Society for the support of E. R. through the International Newton fellowship, grant NF110071, and UKSA for support of STE-QUEST, grant number ST/K006479/1.  We also thank Andrew Hinton, Alexander Niggebaum, Jon Goldwin and Plamen Petrov for useful discussions and critical reading of the manuscript.
\appendix

\section{Full expression of $n_d$ for simultaneous, and sequential imaging.}

\label{AppendixnsTwoLevel}

For fluorescence detection, the expression of the number of scattered photons $n_s$ reaching the photo detector during a time interval $\Delta t$ is given as a function of the atom number $N$ by:
 \begin{eqnarray}
{n_s}(\nu ) = \frac{{\Delta t}}{{h\nu }}\int_V {\int_{ - \infty }^{ + \infty } {\int_{ - \infty }^{ + \infty } I } } (\nu ,\vec r)\sigma \rho (\vec r)g(v,\vec r)dVdvd\nu ,
\label{ndtwolevelFull}
\end{eqnarray}
where $\rho(\vec{r})$ and $g(v,\vec{r})$ are the  atom density and velocity distributions at the position $\vec{r}$ integrated over a volume $V$ \cite{Sorrentino:2013ta}, where $I(\nu,\vec r)$ is the laser intensity at frequency $\nu$ and at the position,  $\vec{r}$, and  $\sigma$ is the scattering cross section, which includes power broadening and doppler shifted detuning.
Assuming a two level atom, $\sigma$ is given by:
 \begin{equation}
\sigma  = {\sigma _0}{\left( {1 + 4\frac{{{{(\Delta  + \nu v/c)}^2}}}{{{\Gamma ^2}}} + \frac{{I(\nu (1 + v/c))}}{{{I_{sat}}}}} \right)^{ - 1}}
\label{sigmaDel}
\end{equation}
with $\Delta$ the detuning with respect to the resonant frequency $\nu_0$ for the atom at rest,  $\Gamma $ the natural line-width (FWHM) of the considered transitions, $v$, the velocity of the atoms relative to the laser source, and $\sigma_0$ is the on-resonant cross section given by the expression  \cite{Steck:moWuU-GK}:
 \begin{equation}
\sigma_0=\frac{ h \nu }{2}\frac{\Gamma}{I_{sat}}.
\end{equation}
The saturation intensity $I_{sat}$ is defined as in \cite{Steck:Rubidium87Data,Steck:moWuU-GK}, and given by the equation:
 \begin{equation}
{I_{sat}} = \frac{{c{\varepsilon _0}{\Gamma ^2}{\hbar ^2}}}{{4{{\left| {\hat \varepsilon  \cdot  \boldsymbol{d}} \right|}^2}}}
\label{eq:satIdef}
\end{equation}
with $\hat \varepsilon$ the unit polarization vector of the light field, and $\boldsymbol{d}$ the atomic dipole moment. The various expressions of $I_{sat}$  for different light polarizations can be found in \cite{Steck:Rubidium87Data}.\\
We assume that the laser source is at rest with respect to the centre of mass of the atom cloud, that there is no atom velocity spread (and so no doppler broadening), and that the laser intensity is constant over the cloud,  $I(\nu,\vec{r})=I$.
Under these conditions, equation (\ref{ndtwolevelFull}) becomes the following one:
 \begin{equation}
{n_s} = N{\gamma _d}\frac{\Gamma }{2}\frac{I}{{{I_{sat}}}}\Delta t .
\label{eq:ndtwolevel}
\end{equation}
Then ${\left[ {{{(\Delta t)}_{min}}} \right]_{2l}}$ of equation (\ref{DtminSimultaneus}) is derived by solving the equation $n_s(\Delta t_{min})\Omega=(n_d)_{min}$ with $(n_d)_{min}$ from section \ref{Section2}.\\
If the atom model is not the simple two level one, but the one represented in figure \ref{TransitionsLosses}, the expression of $n_d$ has to include as well the possibility of atom loss to closed transitions. In this extended model, the scattered photon number $n_s$ is found by solving the following system:
\begin{eqnarray}
\frac{{d{n_s}}}{{dt}} = (N - {n_l})\frac{{I{\sigma _0}}}{{h\nu }}{\gamma _d}\\
\frac{{d{n_l}}}{{dt}} = (N - {n_l})\frac{{I{\sigma _0}}}{{h\nu }}{\gamma _{lost}}
\end{eqnarray}
with $n_l$ the number of atoms lost to the alternative hyperfine transitions, $\gamma_d$ and $\gamma_{lost}$ as defined for equation (\ref{DtMinWithLosses}). 
Because in this work we assume that all the light polarizations are present, and that all the degenerate $m_F$ levels are equally occupied, this multilevel model does not include the optical pumping of the atoms  to different $m_F$ levels by the detection beam.  Also the effect of stimulated emissions to lost transitions has not been modeled since, for small detuning $\Delta$, they represent a second order correction only.\\
For this multilevel model,  $n_s$ as a function of the atom number $N$ is given by the following equation:
 \begin{equation}
{n_s}(t) = \frac{{N{\gamma _d}}}{{{\gamma _{lost}}}}\left( {1 - \exp \left( { - t\frac{I}{{2{I_{sat}}}}\Gamma {\gamma _{lost}}} \right)} \right).
\label{eq:ndmultilevel}
\end{equation}
As for two level system, the minimum detection time $(\Delta t)_{min}$ to reach atom shot noise of equation (\ref{DtMinWithLosses}) is derived by solving the equation $n_s(\Delta t_{min})\Omega=(n_d)_{min}$.\\

\section{Derivation of cross shadowing effect}

\label{AppendixShadow}

We model the effect of section \ref{CrossShadowingSection} by deriving the scattered photon number $n_s$ in the case of cross cloud shadowing.  As in \ref{AppendixnsTwoLevel}, we assume  no atom velocity spread (so no doppler shift) and very narrow laser line width, but the detection beams can have variable light density $I_b$ across $x$, $y$, and $z$ due to the shadowing effect.  As in figure \ref{TwoCloudShadows} the laser beam axis is taken parallel to the $z$ axis, and retro reflected on a mirror on one side of the clouds. 
If $I$ is the incoming beam, for small saturation ratio $I/I_{sat}$, or low optical depth \cite{Pappa:2011ig}, the beam exiting from the first cloud is approximatively given by the Beer Lambert Law:
 \begin{eqnarray}
{I_t}(x,y) = I\exp ( - \sigma {\rho _z}(x,y))\\
 \approx I(1 - \sigma {\rho _z}(x,y))
\end{eqnarray}
with $\rho_z(x,y)$ the atom number density integrated along the $z$ axis. The total intensity of the laser beams illuminating the first $I_{t1}$ and second $I_{t2}$ cloud are:
 \begin{eqnarray}
{I_{t1}} &\approx I\left( {2 - \sigma \left( {{\rho _{1z}}(x,y) + 2{\rho _{2z}}(x,y)} \right)} \right)\\
{I_{t2}} &\approx I\left( {2 - \sigma \left( {2{\rho _{1z}}(x,y) + {\rho _{2z}}(x,y)} \right)} \right),
\end{eqnarray}
with $\rho_{1z(x,y)}$ and $\rho_{2z(x,y)}$ the $z$ integrated atom densities of the first and second cloud at the AI output.
We assume that $\rho_1(x,y,z)=\rho(x,y,z) \xi$ and $\rho_2(x,y,z)=\rho(x,y,z)(1-\xi)$, and that $\rho(x,y,z)$ is normally distributed with standard deviations $\sigma_\rho$ and volume integral $N$, and that the clouds to be at far distance $\gg 2 \sigma_\rho$ from each other. The parameter $\xi$ is the probability that each atom will be in the ground state associated with the first output cloud, and it is related to the AI phase $\phi$ being $\xi=(1+\cos(\phi))/2$. Therefore the scattered photon numbers for the two clouds are:
  \begin{eqnarray}
{n_{1s}} &=& \frac{{\Delta t}}{{h\nu }}\sigma I\xi \int\limits_{ - \infty }^{ + \infty } {\rho (\vec r)\left( {2 - \sigma \rho (\vec r)(2 - \xi )} \right)(1 - \xi {g_\rho }(\vec r))} dV\\
 &= &\frac{{\Delta t}}{{h\nu }}\sigma 2I\xi N\left( {1 - \tau  + \xi (2 - \xi )\Lambda } \right)
\label{eq:ns1shadow}\\
{n_{2s}} &=& \frac{{\Delta t}}{{h\nu }}\sigma I(1 - \xi )\int\limits_{ - \infty }^{ + \infty } {\rho (\vec r)\left( {2 - \sigma \rho (\vec r)(1 + \xi )} \right)(1 - (1 - \xi ){g_\rho }(\vec r))} dV\\
 &= &\frac{{\Delta t}}{{h\nu }}\sigma 2I(1 - \xi )N\left( {1 - \tau  + (1 - \xi )(1 + \xi )\Lambda } \right),
\label{eq:ns2shadow}
\end{eqnarray}
where we included as well the self shadowing of each cloud, and ${g_\rho }(\vec r)$, $\tau$, and $\Lambda$ are given by:
  \begin{eqnarray}
{g_\rho }(\vec r) &=& \int\limits_z^{ + \infty } {\sigma \rho (x,y,\zeta )} d\zeta \\
\tau  &=& \frac{{\sigma N}}{{4\pi \sigma _\rho ^2}} = 2{\rho _{2MAX}}\sigma \\
\Lambda  &=& \frac{{{N^2}{\sigma ^2}}}{{96{\pi ^{5/2}}\sigma _\rho ^5}}.
\end{eqnarray}
We substitute equation (\ref{eq:ns1shadow}) and  (\ref{eq:ns2shadow}) in the definition of $\zeta_m$, and $\delta_\zeta$ obtaining the expression of the shadowing error induced on the population estimate of the AI output clouds  of equation \ref{eq:errorFromShadowingNormalized}.
\section{Branching ratio}

\label{AppendixBranchingRatio}

The decay branching ratio from an hyperfine level $F'$ of an exited state $J'$ to an hyperfine level $F$ of a ground state $J$ is given by the following equation as a function of Wigner $3-j$,and $6-j$ symbols:
 
\begin{eqnarray}
  \rm{br}_{F’ \to F}  = \frac{{{\Gamma _{F \to F'}}}}{{\sum\limits_F {{\Gamma _{F \to F'}}} }} \label{branchingdef} \nonumber \\ 
   = \left( {2F + 1} \right)\left( {2J' + 1} \right){\left\{ {\begin{array}{*{20}{c}}
  J&{J'}&1 \\ 
  {F'}&F&I 
\end{array}} \right\}^2}\cdot& \sum\limits_{q,m{'_F}} {{{\left( {\begin{array}{*{20}{c}}
  {F'}&1&F \\ 
  {m{'_F}}&q&{ - {m_F}} 
\end{array}} \right)}^2}} ,
\label{branching}
\end{eqnarray}

where we sum over the possible polarization of the light ($q=-1,0,1$) and the hyperfine $m{'_F}$ sub levels of $F’$, and $m_F$ is equal to $m'{_F}+q$. The previous equation can be found from the expression of the decay rate from an hyperfine level $F'$ of an exited state $J'$ to an hyperfine level $F$ of a ground state $J$  \cite{Loudon:2000up, Steck:moWuU-GK}:
 \begin{equation}
{\Gamma _{F \to F'}} = \sum\limits_{q,m{'_F}} {\frac{{\omega _0^3}}{{3\pi {\varepsilon _0}\hbar {c^3}}}{{\left| {\left\langle {F\left( {m{'_F} + q} \right)\left| {e{r_q}} \right|F'm{'_F}} \right\rangle } \right|}^2}} .
\label{GammaFFp}
\end{equation} 
By using the Wigner-Eckart theorem \cite{Brink:1962uy,Steck:moWuU-GK} in the previous expression and then by summing ${\Gamma _{F \to F'}}$ over all the final ground levels $F$ in equation (\ref{branchingdef}), equation (\ref{branching}) is derived.  For completeness in table \ref{BranchingRatioTable} we present the branching ratio to closed transition for some Alkali atoms, and see how they are very similar across different elements. 
\begin{table}
\caption{\label{BranchingRatioTable}Branching ratio to closed transition for various Alkali elements. $^{7}\rm{Li}$,$^{23}\rm{Na}$,$^{39}\rm{K}$, and $^{41}\rm{K}$ have the same branching ratio as $^{87}\rm{Rb}$.}
\begin{indented}
\item[]\begin{tabular}{@{}lll}
\br
 Element&Transition&$\rm{br}$\\
\mr
 $^{87}\rm{Rb}$& $F'=2 \to F=1$ &  50\%\\
 $^{87}\rm{Rb}$&  $F'=1 \to F=1$&83.3\%\\
 $^{85}\rm{Rb}$& $F'=3 \to F=2$&44.4\%\\
 $^{85}\rm{Rb}$&$F'=2 \to F=2$&77.8\%\\
\br
\end{tabular}
\begin{tabular}{@{}lll}
\br
 Element&Transition&$\rm{br}$\\
\mr
$^{6}\rm{Li}$&$F'=3/2 \to F=1/2$&  55.6\% \\
$^{6}\rm{Li}$&$F'=1/2 \to F=1/2$& 83.9\%\\
$^{133}\rm{Cs}$&$F'=4 \to F=3$ &41.7\%\\
$^{133}\rm{Cs}$&$F'=3 \to F=3$&75 \%\\
\br
\end{tabular}
\end{indented}
\end{table}
\bibliographystyle{iopart-num}
\section*{References}
\bibliography{references}

\end{document}